\documentclass{aa}

\usepackage{graphicx}
\usepackage{xcolor}

\usepackage[normalem]{ulem}
\usepackage[varg]{txfonts}
\usepackage[stable]{footmisc}
\usepackage{url}
\begin{document} 

    \title{Astrochemical relevance of VUV ionization of large PAH cations\thanks{The dataset associated with this work can be found under 10.5281/zenodo.3899775}}
   \author{G. Wenzel\inst{1} \and
           C. Joblin\inst{1}\fnmsep\thanks{Corresponding author.} \and
           A. Giuliani\inst{2,3} \and
           S. Rodriguez Castillo\inst{1,4} \and
           G. Mulas\inst{1,5} \and
           M. Ji\inst{1} \and
           H. Sabbah\inst{1,6} \and
           S. Quiroga\inst{7} \and
           D. Pe\~na\inst{7} \and 
           L. Nahon\inst{2}
     }

   \institute{Institut de Recherche en Astrophysique et Plan\'{e}tologie (IRAP),                 Universit\'{e} de Toulouse (UPS), CNRS, CNES, 9 Avenue du Colonel Roche,            F-31028 Toulouse, France \\ 
             \email{christine.joblin@irap.omp.eu}
          \and 
             Synchrotron SOLEIL, L'Orme des Merisiers, F-91192 Saint Aubin, Gif-sur-Yvette, France
         \and
              INRAE, UAR1008, Transform Department, Rue de la G\'eraudi\`ere, BP 71627,
              F-44316 Nantes, France
         \and
             Laboratoire de Chimie et Physique Quantiques (LCPQ/IRSAMC), Universit\'{e} de Toulouse (UPS), CNRS, 118 Route de Narbonne, F-31062 Toulouse, France
         \and
             Istituto Nazionale di Astrofisica -- Osservatorio Astronomico di Cagliari, Via della Scienza 5, I-09047 Selargius (CA), Italy
         \and
             Laboratoire Collisions Agr\'{e}gats R\'{e}activit\'{e} (LCAR/IRSAMC), Universit\'{e} de Toulouse (UPS), CNRS, 118 Route de Narbonne, F-31062 Toulouse, France
         \and  
         Centro de Investigaci\'on en Qu\'imica Biol\'oxica e Materiais Moleculares (CiQUS) and Departamento de Qu\'imica Org\'anica, Universidade de Santiago de Compostela, E-15782 Santiago de Compostela, Spain
         \\
             }

   \date{Received 00 Month 2020 / Accepted 00 Month 2020}

% \abstract{}{}{}{}{} 
% 5 {} token are mandatory
 
  \abstract
  % context heading (optional), leave it empty if necessary 
   {As a part of interstellar dust, polycyclic aromatic hydrocarbons (PAHs) are processed by the interaction with vacuum ultraviolet (VUV) photons that are emitted by hot young stars. This interaction leads to the emission of the well-known aromatic infrared bands but also of electrons, which can significantly contribute to the heating of the interstellar gas.}
   {Our aim is to investigate the impact of molecular size on the photoionization properties of cationic PAHs.}
   {Trapped PAH cations of sizes between 30 and 48 carbon atoms were submitted to VUV photons in the range of 9 to 20\,eV from the DESIRS beamline at the synchrotron SOLEIL. All resulting photoproducts including dications and fragment cations were mass-analyzed and recorded as a function of photon energy.}
   {Photoionization is found to be predominant over dissociation at all energies, which differs from an earlier study on smaller PAHs. The photoionization branching ratio reaches  0.98 at 20\,eV for the largest studied PAH. The photoionization threshold is observed to be between 9.1 and 10.2\,eV, in agreement with the evolution of the ionization potential with size. Ionization cross sections were indirectly obtained and photoionization yields extracted from their ratio with theoretical photoabsorption cross sections, which were calculated using time-dependent density functional theory. An analytical function was derived to calculate this yield for a given molecular size.}
   {Large PAH cations could be efficiently ionized in \ion{H}{I} regions and provide a contribution to the heating of the gas by photoelectric effect. Also, at the border of or in \ion{H}{II} regions, PAHs could be exposed to photons of energy higher than 13.6\,eV. Our work provides recipes to be used in astronomical models to quantify these points.}

    
   \maketitle
%
%-------------------------------------------------------------------

\section{Introduction}

Polycyclic aromatic hydrocarbons (PAHs) play a major role in the physics and chemistry of photodissociation regions (PDRs). They strongly absorb vacuum ultraviolet (VUV) photons that are emitted by hot young stars and relax by emission in the aromatic infrared bands (AIBs). The interaction with VUV photons can lead to other relaxation processes including ionization and dissociation. All these processes together with reactive processes involving in particular electrons and hydrogen (H, H$_2$) govern the evolution of the PAH population in the diffuse interstellar medium \citep{lepage2003}, in circumstellar disks \citep{visser2007}, and in reflection nebulae \citep{montillaud2013}. The results of these chemical models suggest that large PAHs with a typical carbon number, $N_\mathrm{C}$, of 50 or more dominate the AIB emission which led to the grandPAH hypothesis that large and possibly compact PAHs dominate the emission in bright PDRs \citep{andrews2015}. In some regions associated with these PDRs, large PAHs are expected to be ionized reaching even the dicationic stage \citep{tielens2005, andrews2016}.

In a previous study we have investigated the branching ratio between ionization and fragmentation upon VUV irradiation for medium-sized PAH cations \citep{zhen2016a} with an $N_\mathrm{C}$ between 16 and 24. For all of these cations, fragmentation was observed to be the dominant channel at least up to a photon energy of $13.6\,\mathrm{eV}$ which is relevant for \ion{H}{I} regions. In the case of larger PAH cations, ionization is expected to be by far the dominant channel as suggested by the study of the hexa-peri-hexabenzocoronene (HBC) cation, C$_{42}$H$_{18}^+$, by \cite{zhen2015}. Here, our objective is to quantify the growing importance of ionization as the molecular size increases. Following \citet{zhen2016a}, we have studied the photoprocessing of PAH cations with an $N_\mathrm{C}$ between 30 and 48 atoms over the $9.5-20.0\,\mathrm{eV}$ VUV range. Photon energies above the Lyman limit are relevant to PAHs observed at the border of ionization fronts in PDRs \citep{vicente2013}, as well as in \ion{H}{II} regions \citep{compiegne2007}.

Although we report here also the branching ratio between ionization and dissociation, our analysis is focused on ionization. More specifically, we derive the photoionization yield, which is important to model the charge balance of PAHs and its impact on the AIB spectrum \citep{bakes2001a}, but also to evaluate the contribution of these species to the photoelectric heating rate \citep{bakes1994,weingartner2001b}. The experimental method is described in Sect.~\ref{sec:exp} and the results are presented in Sect.~\ref{sec:res}. In Sect.~\ref{sec:astro}, we discuss the astrophysical implications and propose recipes to be used in astrophysical models. We conclude in Sect.~\ref{sec:con}.

\section{Experimental method and data analysis}\label{sec:exp}

\begin{figure}
\resizebox{\hsize}{!}{
    \centering
    \includegraphics{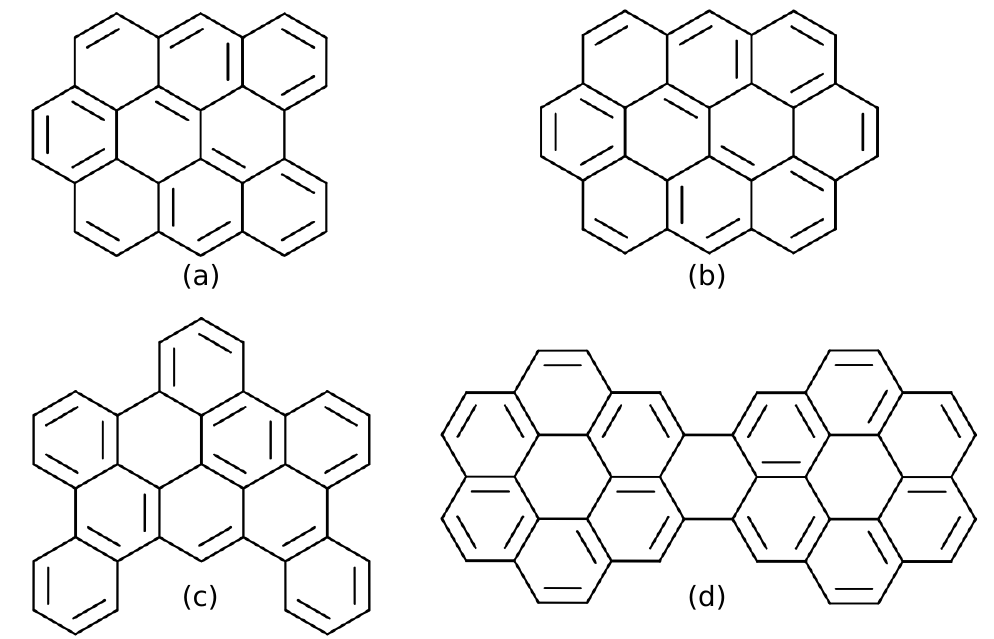}}
    \caption{Molecular structures of the studied PAHs, namely (a) benzobisanthene, C$_{30}$H$_{14}^+$, (b) ovalene, C$_{32}$H$_{14}^+$, (c) DBPP, C$_{36}$H$_{18}^+$, and (d) dicoronylene, C$_{48}$H$_{20}^+$.}
    \label{fig:molstruc}
\end{figure}

\begin{figure*}[htbp]
    \centering
    \includegraphics[width=\textwidth]{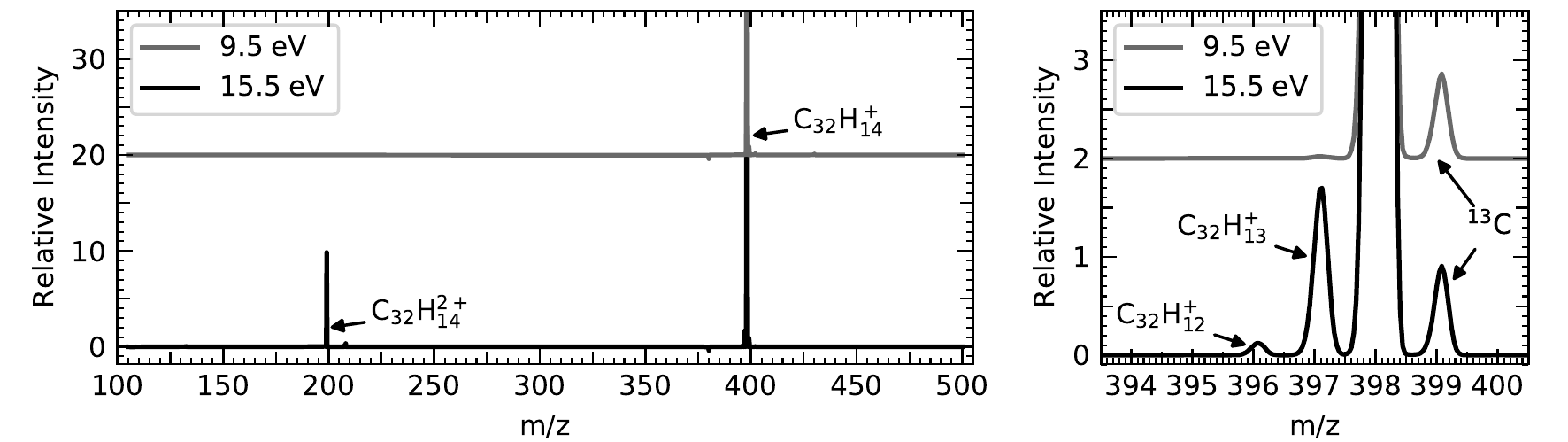}
    \caption{Mass spectra of the ovalene parent cation, C$_{32}$H$_{14}^+$, $m/z=398$, at two different photon energies. At $9.5\,\mathrm{eV}$, none of the photoionization or photodissociation channels are opened whereas at $15.5\,\mathrm{eV}$, the doubly ionized parent ion, C$_{32}$H$_{14}^{2+}$, at $m/z=199$ is observed as well as the photofragments due to the loss of one and two hydrogen atoms, C$_{32}$H$_{13}^{+}$ and C$_{32}$H$_{12}^{+}$ at $m/z=397$ and $396$, respectively.}
    \label{fig:ms}
\end{figure*}

We have used the Thermo Scientific\texttrademark\ LTQ XL\texttrademark\ linear ion trap (LTQ ion trap) as described in \cite{milosavljevic2012}, which is available at the VUV beamline DESIRS at the synchrotron SOLEIL \citep{nahon2012}.

The production of PAH cations in the LTQ ion trap was performed using an atmospheric pressure photoionization (APPI) source which required the species of interest to be in solution before their injection with a syringe. This part was a major limitation on the size range of PAHs we could study due to the non-solubility of large PAHs. Four large PAH cations with $N_\mathrm{C}$ ranging from 30 to 48 could be investigated in this study, namely (a) benzobisanthene, C$_{30}$H$_{14}^+$, (b) ovalene, C$_{32}$H$_{14}^+$, (c) dibenzophenanthropentaphene (DBPP), C$_{36}$H$_{18}^+$, and (d) dicoronylene, C$_{48}$H$_{20}^+$. Sample (b) originated from Janssen Chimica (Belgium), samples (a) and (c) from the PAH Research Institute in Greifenberg (Dr. Werner Schmidt). The synthesis of compound (d) is briefly reported in Appendix~\ref{app:dico}. The molecular structures of the studied species are depicted in Fig.~\ref{fig:molstruc}.
Emptying the syringe was performed at a flow rate which was kept constant for each experiment.
The used flow rate was 4$\,\mu\mathrm{l\,min}^{-1}$ for compounds (a) and (b), $6\,\mu\mathrm{l\,min}^{-1}$ for compound (c), and 10$\,\mu\mathrm{l\,min}^{-1}$ for compound (d).
The presence of UV irradiation from a Kr discharge lamp ensured a soft creation of PAH cations without fragmentation \citep{giuliani2012}. The formed cations were then guided through ion optics into the LTQ ion trap in which a constant He pressure of $p \approx 10^{-3}\,\mathrm{mbar}$ was held. The ions were cooled by the collisions with He atoms and the PAH cations of interest, the so-called parent ions, were isolated through specific mass selection and ejection of other species from the ion trap including the \element[ ][13]{C} isotopomers.

The parent ions were then submitted to the VUV synchrotron radiation which was tuned from $9.5$ to $20.0\,\mathrm{eV}$ in steps of $0.1$, $0.2$, $0.3$, or $0.5\,\mathrm{eV}$ depending on the photon energy range, with the exception of C$_{30}$H$_{14}^+$, for which we were able to scan only at low energies up to $15.5\,\mathrm{eV}$. Higher harmonics of the VUV undulator synchrotron radiation with photon energies lower than $16.0\,\mathrm{eV}$ were filtered out by a gas filter filled with Ar gas to a pressure of $0.23\,\mathrm{mbar}$. Above $16.0\,\mathrm{eV}$ no such gas filtering is necessary. The photon flux was measured with an IRD AXUV100 calibrated Si photodiode for a monochromator exit slit width of $200\,\mathrm{\mu m}$ and was between $0.8$ and $2.8\,10^{12}\,\mathrm{photons\,s^{-1}}$ over the studied photon energy range. A typical photon flux can be derived using a previous calibration of the beam size as a function of photon energy \citep{douix2017}, yielding values of $1.5 - 5.2\,10^{14}\,\mathrm{photons\,cm^{-2}s^{-1}}$. In order to limit possible two photon consecutive absorption processes, we tuned the photon flux by changing (i) the irradiation time from $0.8$ to $0.2\,\mathrm{s}$ for the lower and higher photon energy ranges, respectively, and (ii) the monochromator exit slit width from $200\,\mu \mathrm{m}$ at low energies to $70\,\mu \mathrm{m}$ at high energies, except for dicoronylene for which values of $400$ and $100\,\mu \mathrm{m}$ at low and high energies, respectively, were used to improve the signal-to-noise ratio. 
The photon dose was assumed to be linearly proportional to both the irradiation time and the monochromator exit slit width.
The probability of two photon absorption processes could be estimated on the formation of triply charged parent ions, yielding only very small relative intensities below 2\,\% of the total number of photoproducts.

Depending on the acquisition time, a few hundred mass spectra were recorded at each photon energy and averaged to yield one mass spectrum per photon energy step. Following the same procedure, we also recorded blank mass spectra at each photon energy by selecting a mass close to but different enough from each parent ion. This allowed us to perform background subtraction which eliminates contamination peaks from the mass spectra. The averaging procedure provides us with a statistical standard error (see Appendix~\ref{app:error}). As an example, the background subtracted mass spectra for ovalene, which has a mass-to-charge ratio of $m/z = 398$, are depicted in Fig.~\ref{fig:ms} at two different photon energies of $9.5\,\mathrm{eV}$ and $15.5\,\mathrm{eV}$. The parent ion, C$_{32}$H$_{14}^+$, is well isolated, the \element[][13]{C} isotopic parent ion has a residual contribution of less than 1\,\% remaining in the ion trap. By increasing the photon energy, three different secondary ions can be observed and unambiguously separated, namely the H and 2H/H$_2$ loss, and the main doubly ionized parent ion channels. For the presented example of the ovalene cation, C$_{32}$H$_{14}^+$, these species are C$_{32}$H$_{13}^+$, C$_{32}$H$_{12}^+$, and C$_{32}$H$_{14}^{2+}$, respectively (see Fig.~\ref{fig:ms}). When extracting the peak intensities as will be done in the following, one has to consider the detector gain efficiency that varies with the charge and the mass of the ions of interest. Recommended scaling factors were therefore applied (see Appendix~\ref{app:actionspec}).

\section{Results and discussion}\label{sec:res}

\subsection{Action spectra and branching ratio}\label{sec:act}

\begin{figure*}[htbp]
    \centering
    \includegraphics[width=\textwidth]{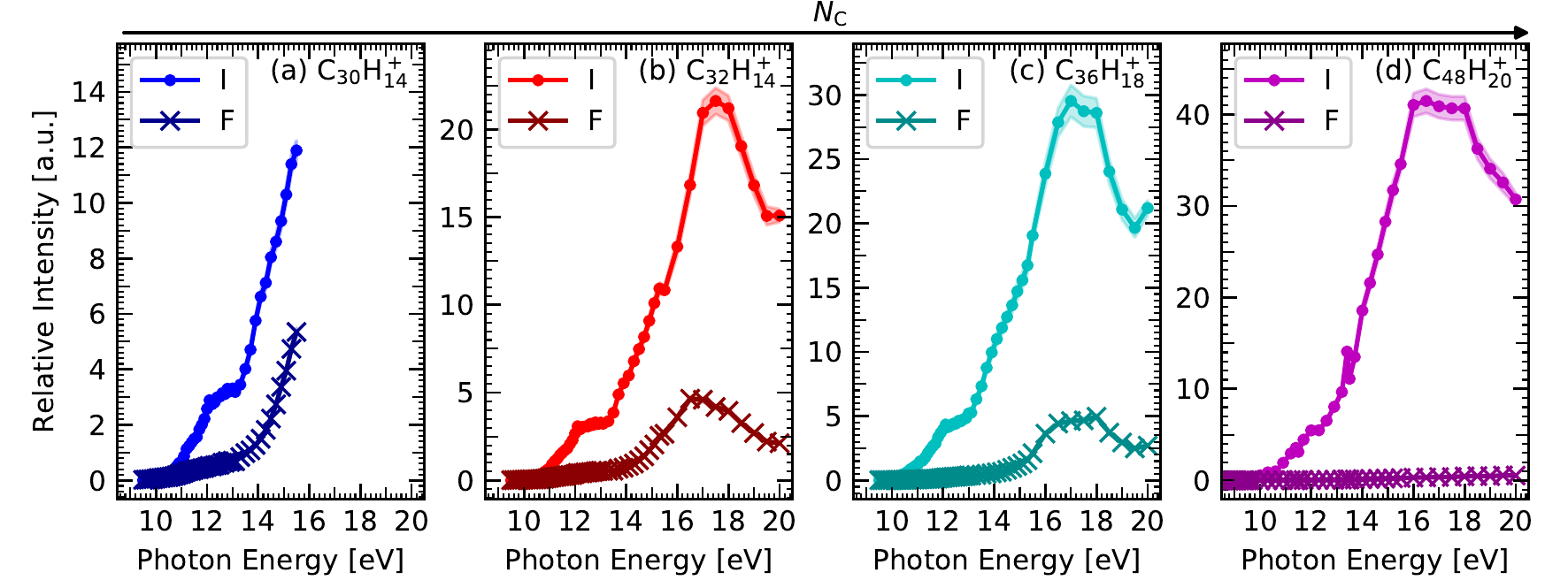} \caption{Action spectra of the photoproducts, dications (I) and fragments (F), as a function of photon energy for all studied PAH cations, (a) benzobisanthene, (b) ovalene, (c) DBPP, and (d) dicoronylene after absorption of a VUV photon. Relative intensities as explained in Appendix~\ref{app:actionspec}.}
    \label{fig:ri}
\end{figure*}

\begin{figure}[htbp]
    \resizebox{\hsize}{!}{
    \includegraphics{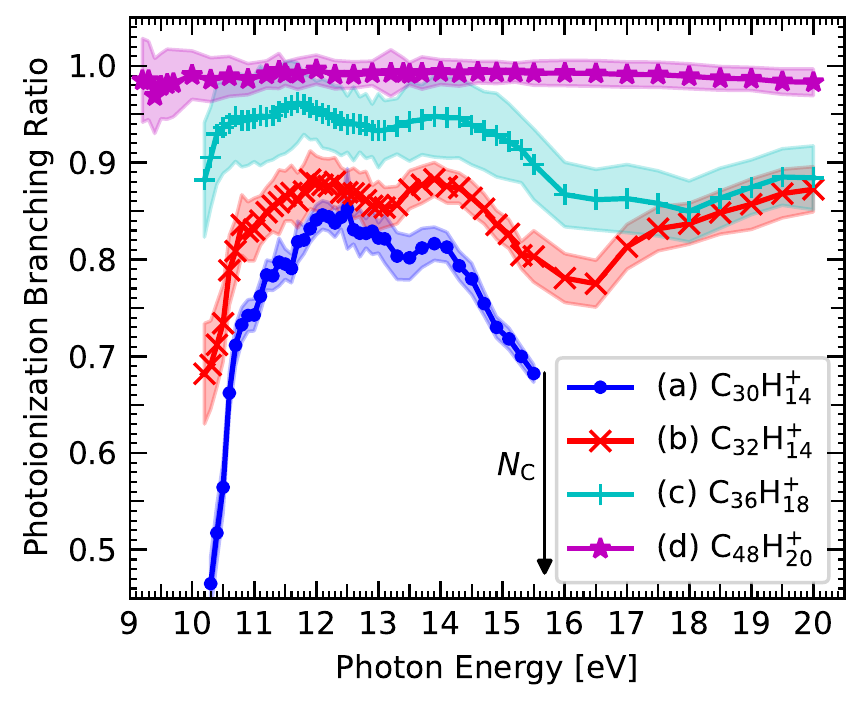}}
    \caption{Photoionization branching ratio relative to photodissociation as a function of photon energy, for (a) benzobisanthene, C$_{30}$H$_{14}^+$, (b) ovalene, C$_{32}$H$_{14}^+$, (c) DBPP, C$_{36}$H$_{18}^+$, and (d) dicoronylene, C$_{48}$H$_{20}^+$, after absorption of a single VUV photon.}
    \label{fig:br}
 \end{figure}

\begin{figure*}[htbp]
    \centering
    \includegraphics[width=\textwidth]{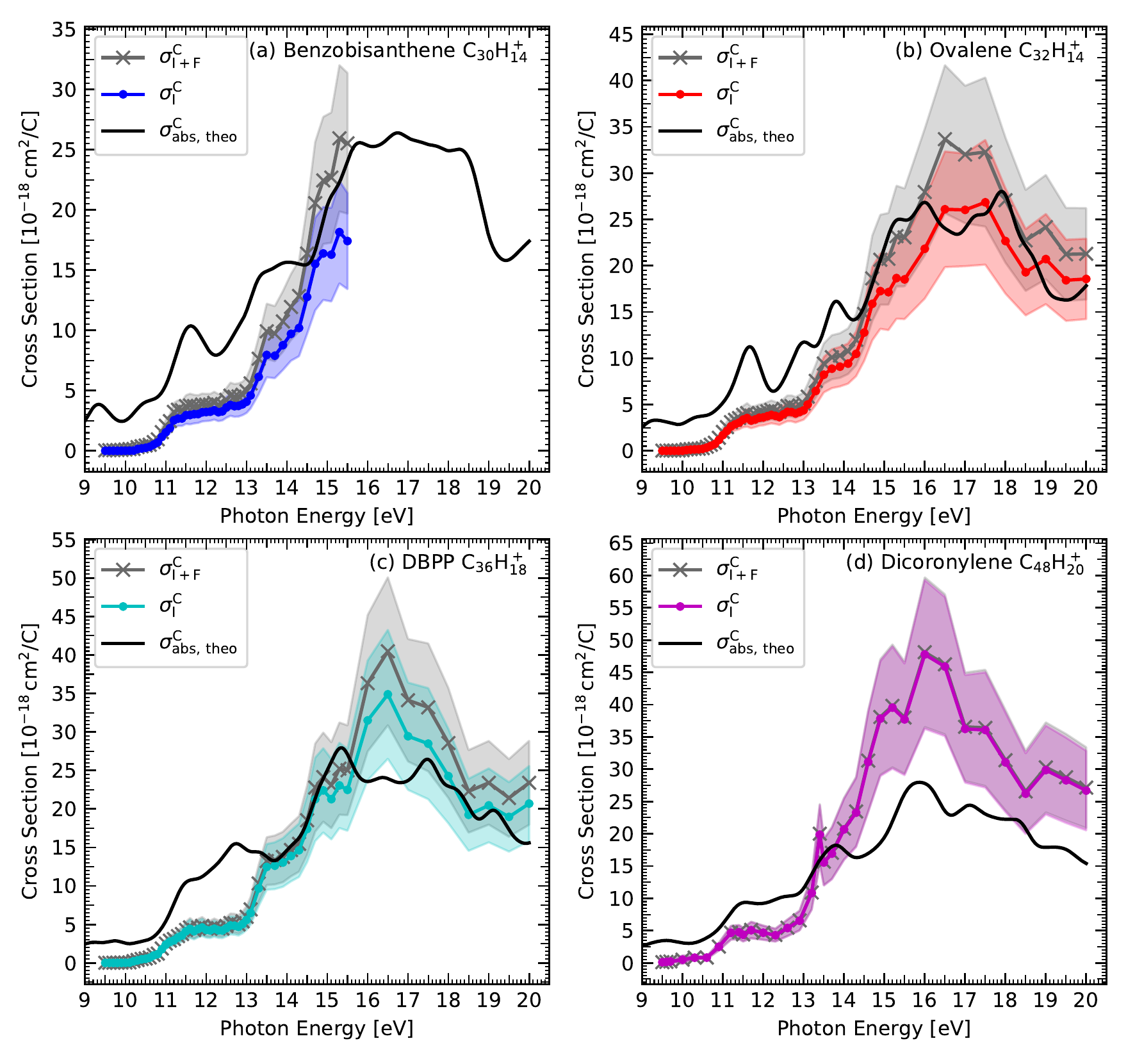}
    \caption{Experimentally obtained total action and ionization cross sections per C atom, $\sigma_\mathrm{I+F}^\mathrm{C}$ and $\sigma_\mathrm{I}^\mathrm{C}$, respectively, of the studied PAH cations as a function of photon energy, for (a) benzobisanthene, (b) ovalene, (c) DBPP, and (d) dicoronylene after absorption of a single VUV photon, compared to their by TD-DFT computed theoretical photoabsorption cross sections per C atom, $\sigma_\mathrm{abs,\,theo}^\mathrm{C}$.}
    \label{fig:cs}
\end{figure*}

The action spectra are determined following the procedure described in Appendix~\ref{app:actionspec} yielding relative intensities of the photoproducts as used in previous work \citep{zhen2016a}. The resulting spectra for the photoionization (dication, denoted I) and photodissociation (fragments, denoted F) channels of the four studied PAH cations are shown in Fig.~\ref{fig:ri} as a function of the photon energy. The F channel remains small for all investigated PAH cations at all photon energies and is barely notable for the dicoronylene cation in Fig.~\ref{fig:ri}. More specifically, Fig.~\ref{fig:br} shows that the branching ratio (BR) for photoionization relative to photodissociation increases significantly with increasing $N_\mathrm{C}$ and reaches a minimal value of 0.98 for the largest studied cation. This trend of large PAHs differs from what was observed in our earlier study of medium-sized PAH cations for which a larger fraction of fragments was observed \citep{zhen2016a}. It is in line with the ionization BR of about 0.97 at 20\,eV which was derived by \cite{zhen2015} for the HBC cation, C$_{42}$H$_{18}^+$, by operating their home-made ion trap setup at the DESIRS beamline. The authors also reported a value of $0.70 \pm 0.10$ for the ionization BR of the ovalene cation at 20\,eV, which can be compared to a value of $0.87 \pm 0.02$ in our experiment. This difference can be interpreted by the low mass resolution achieved in the former experiments which impacted both the isolation of the $^{12}$C parent isotopomer before irradiation and the quantification of the abundance of --H fragments in the photoproducts.

\begin{table}[htbp]
\caption{\label{table:IP}For doubly ionized PAH cations, $\mathrm{PAH}^+ \xrightarrow{} \mathrm{PAH}^{2+}$, values of the theoretical adiabatic ionization potentials, $\mathrm{IP}^{2+}$, and measured appearance energies, $\mathrm{AE}^{2+}$. We also list here our recorded $\mathrm{AE}^{3+}$ as obtained from the ionization of $\mathrm{PAH}^{2+} \xrightarrow{} \mathrm{PAH}^{3+}$.}
\centering
\begin{tabular}{cccc}
\hline\hline
\multicolumn{1}{c}{\textbf{Formula}} & \multicolumn{1}{c}{\textbf{IP$\boldsymbol{^{2+}}$ [eV]}} & 
\multicolumn{1}{c}{\textbf{AE$\boldsymbol{^{2+}}$ [eV]}} & \multicolumn{1}{c}{\textbf{AE$\boldsymbol{^{3+}}$ [eV]}} \\
\hline 
$\mathrm{C}_{30}\mathrm{H}_{14}^+$    & 9.66\tablefootmark{a} & $10.2 \pm 0.1$ & $14.1 \pm 0.2$ \\
$\mathrm{C}_{32}\mathrm{H}_{14}^+$    & 9.82\tablefootmark{a} & $10.0 \pm 0.1$ & $13.9 \pm 0.2$ \\
$\mathrm{C}_{36}\mathrm{H}_{18}^+$    & 9.94\tablefootmark{b} & $10.0 \pm 0.1$ & $13.9 \pm 0.2$ \\
$\mathrm{C}_{48}\mathrm{H}_{20}^+$    & 8.84\tablefootmark{a} & $9.1 \pm 0.2$  & $-$\tablefootmark{c} \\
\hline
\end{tabular}
\tablefoot{\\\tablefoottext{a}{Taken from \citet{malloci2007db}.}\\
\tablefoottext{b}{Calculated for this work according to \citet{malloci2004}.}\\
\tablefoottext{c}{Trication peak out of mass range.}
}
\end{table}

From Figure~\ref{fig:ri}, we derived appearance energies for the formation of PAH$^{2+}$ from PAH$^+$, $\mathrm{AE}^{2+}$. The values are listed in Table~\ref{table:IP} and compared to the corresponding computed values for the adiabatic ionization potentials, $\mathrm{IP}^{2+}$, which are extracted from the Theoretical Spectral Database of PAHs\footnote{http://astrochemistry.oa-cagliari.inaf.it/database/} \citep{malloci2007db} or calculated at the same level of theory for the missing $\mathrm{IP}^{2+}$ of the DBPP cation according to \cite{malloci2007}. Experimental and theoretical values are found to be in good agreement considering the accuracy of $\sim$0.3\,eV for the calculated values. The trend of a slow decrease of $\mathrm{IP}^{2+}$ with $N_\mathrm{C} \gtrsim 30$ carbon atoms reported by \cite{malloci2007} is confirmed.

\subsection{Photoionization cross sections}\label{sec:PIcross}

Experimental total action cross sections per carbon atom, $\sigma^\mathrm{C}_\mathrm{I+F}$, were obtained following the procedure described in Appendix~\ref{app:cs}. The photoionization cross sections, $\sigma^\mathrm{C}_\mathrm{I}$, were then derived by using the branching ratio depicted in Fig.~\ref{fig:br}. The $\sigma^\mathrm{C}_\mathrm{I+F}$ curves are expected to provide a lower value for the absolute photoabsorption cross sections, $\sigma^\mathrm{C}_\mathrm{abs}$ (see Eq.~(\ref{eq:sabs})). Since $\sigma^\mathrm{C}_\mathrm{abs}$ of the studied cations could not be extracted from the performed experiment and have not been reported so far in the literature, we compare these curves with the theoretical photoabsorption cross sections, $\sigma^\mathrm{C}_\mathrm{abs,\,theo}$, which have been computed using Time-Dependent Density Functional Theory (TD-DFT) in line with our previous work \citep{malloci2004,malloci2007db} and as described in Appendix~\ref{app:thcalcs}. All obtained cross sections, experimental and theoretical, are displayed in Fig.~\ref{fig:cs} and compared to each other in the following at high ($>14$\,eV) and low ($<14$\,eV) energies.

\begin{figure}[htbp]
    \resizebox{\hsize}{!}{
    \includegraphics{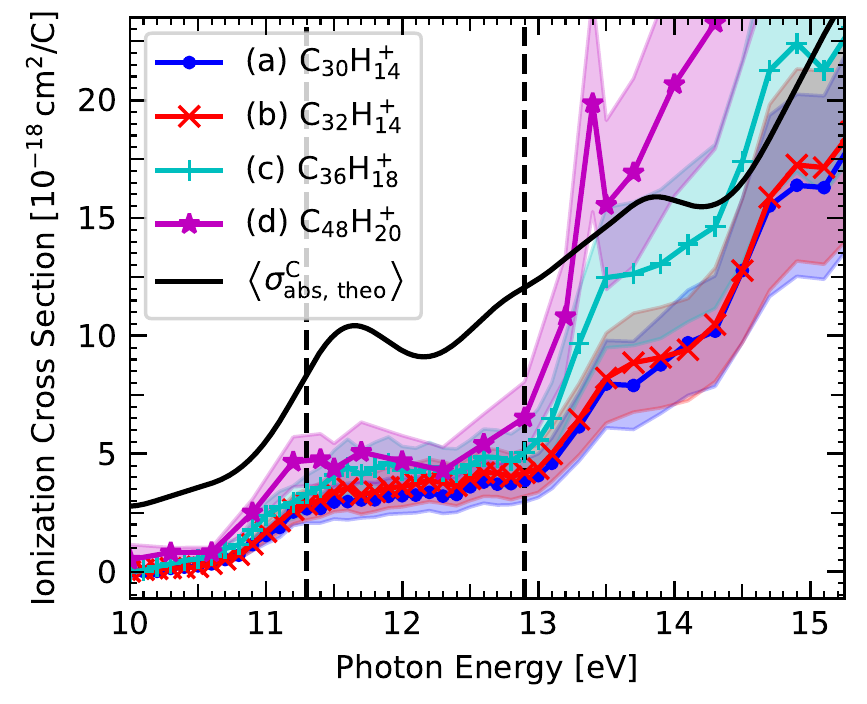}}
    \caption{Experimentally obtained photoionization cross sections per C atom, $\sigma_\mathrm{I}^\mathrm{C}$, of the studied PAH cations as a function of photon energy, (a) benzobisanthene, (b) ovalene, (c) DBPP, and (d) dicoronylene after absorption of a single VUV photon. The average of the calculated photoabsorption cross sections, $\left<\sigma_\mathrm{abs,\,theo}^\mathrm{C}\right>$, is also presented. The dashed vertical lines mark the transitions between the different ionization regimes.}         
    \label{fig:ics}
\end{figure}

Above 14\,eV, the cross sections are globally consistent (see Fig.~\ref{fig:cs}). Still, the values of $\sigma^\mathrm{C}_\mathrm{I+F}$ are found to be systematically larger than those of $\sigma^\mathrm{C}_\mathrm{abs,\,theo}$ around the peak at 17\,eV. In addition, there is a trend of increasing $\sigma^\mathrm{C}_\mathrm{I+F}$ at the peak with molecular size. The case of C$_{48}$H$_{20}^+$ has to be considered with caution though due to a less accurate calibration procedure (see Appendix~\ref{app:cs}). On the contrary to the experimental cross sections, the values of $\sigma^\mathrm{C}_\mathrm{abs,\,theo}$ stay close to each other, which is expected from the proportionality of the photoabsorption cross sections with $N_\mathrm{C}$. Still, it is not yet possible to access how precise the calculated cross sections are. The comparison with an experimental photoabsorption cross section at high energies (up to 30\,eV) has been done so far only for neutral anthracene, C$_{14}$H$_{10}$ \citep{malloci2004}. It is interesting to mention that this comparison reveals an overall good agreement between the calculated and measured cross sections but with differences on the band positions and widths (in the theoretical spectra the band width is artificial). Also around the high energy peak observed at 18\,eV, the discrepancy appears similar to the one illustrated in Fig.~\ref{fig:cs} in the case of C$_{32}$H$_{14}^+$ and C$_{36}$H$_{18}^+$.

Below 14\,eV, $\sigma^\mathrm{C}_\mathrm{I+F}$ and $\sigma^\mathrm{C}_\mathrm{I}$ are comparable since the fragmentation is negligible. A plateau is observed from 11.3 to 12.9\,eV in these experimental cross sections whereas the theoretical photoabsorption cross section exhibits strong bands (see Fig.~\ref{fig:cs}).  There may be a number of possible reasons for the absence of strong, discrete absorption bands in $\sigma^\mathrm{C}_\mathrm{I}$. Ionization cross sections in PAHs may display considerable structure, corresponding both to autoionization resonances and to the opening of channels as energy increases, making accessible additional excited states in the resulting ion with one electron less, i.e., vertical transitions \citep[see e.g.][]{brechignac2014}.

Theoretical considerations have shown that the orbital picture of ionization involving valence one-electron bands is severely contaminated by shake-up states, which involve two electrons, one promoted to an excited bound state and the other to the ionization continuum \citep{deleuze2001}. The authors have shown that, in the case of $\pi$ orbitals, this happens at energies as low as 8\,eV for the first ionization (case of neutral PAHs).
The density of excited states becomes quickly very large, blending in a quasi-continuum as energy increases \citep{deleuze2001}. In addition, each electronic excited state may display vibrational structure on top of it, i.e., vibronic states, producing further structure on a smaller scale. One thus expects, qualitatively, a jump in the ionization cross section when a major channel becomes accessible via a valence one-electron transition, followed by a long plateau-like tail produced by all states coupling to it due to electronic correlation and vibronic coupling.
On top of this, there may be electronic transitions to excited states of the parent molecule which have low coupling with the ionization continuum, therefore preferentially relaxing via radiationless transitions to lower electronic states. These do not show in the ionization cross sections. The theoretical method used here to compute the absolute, total cross section does not distinguish among different categories of electronic excitation, they are in principle all included together (except vibronic coupling) and cannot be distinguished in these calculations. More complex techniques involving many-body theory can compute the structure of the ionization cross sections with considerable accuracy \citep[see e.g.][for a recent review]{baiardi2017}. However, their computational cost would be extremely high for the species considered here, and they are out of the scope of the present work, in which such detailed structure is not resolved anyway.

Below 14\,eV, it is clear that there is a part of the photoabsorption cross section that does not lead to ionization, which we referred to as $\sigma^\mathrm{'}$ in Eq.~(\ref{eq:sabs}). These correspond to excitations which involve fast relaxation via a strong coupling between electronic states and with nuclear states. This leads to vibrational excitation of the parent ion, which is expected to subsequently relax its energy by radiative cooling since no fragmentation is observed. At energies higher than 14\,eV, evidence for such transitions, if they exist, is hindered by the precision of our experimental and theoretical data, as discussed above.

\section{Astrophysical recipes}\label{sec:astro}

\subsection{Charge state of astro-PAHs}
\begin{figure}[htbp]
    \resizebox{\hsize}{!}{
    \includegraphics{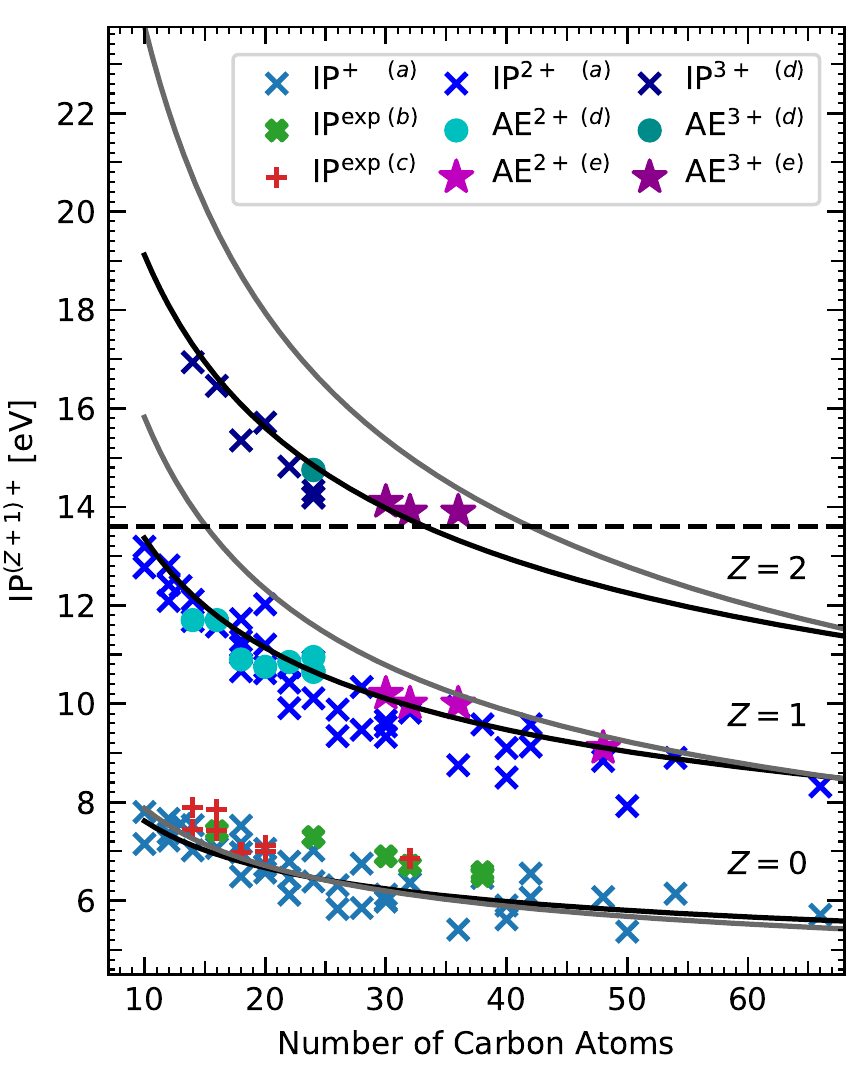}}
    \caption{Theoretically calculated ionization potentials, IP$^{(Z+1)+}$, experimentally obtained IP$^\mathrm{exp}$ and appearance energies, AE$^{(Z+1)+}$, as a function of number of carbon atoms. For $Z=0,1,2$, the values refer to the transition from $\mathrm{PAH}^{(Z)+} \xrightarrow{} \mathrm{PAH}^{(Z+1)+}$. The black and gray curves are the modeled IP$^{(Z+1)+}_{WD}$ from \citet{weingartner2001a} and IP$^{(Z+1)+}_{BT}$ from \citet{bakes1994a}, respectively, as an estimate of the IP$^{(Z+1)+}$ evolution as a function of PAH size and charge, $Z$ (see Eqs.~(\ref{eq:IP}) and (\ref{eq:IP_BT}) and text for details). The dashed horizontal line marks the $13.6\,\mathrm{eV}$ photon energy cut-off for \ion{H}{I} regions.}
    \tablefoot{\\\tablefoottext{a}{Taken from \citet{malloci2007}.}\\
    \tablefoottext{b}{Taken from \citet{clar1981}.}\\
    \tablefoottext{c}{Taken from \citet{hager1988}.}\\
    \tablefoottext{d}{Taken from \citet{zhen2016a}.}\\
    \tablefoottext{e}{This work.}
    }
    \label{fig:IP_AE}
\end{figure}

A couple of modeling studies on the charge state of astro-PAHs have considered that these species could reach the dication and marginally the trication states \citep{bakes2001a, weingartner2001b, andrews2016}. In Figure~\ref{fig:IP_AE}, we compiled the ionization potentials (IP$^{(Z+1)+}$) from neutral to cation ($Z=0$), cation to dication ($Z=1$), and dication to trication ($Z=2$), which have been obtained from calculations \citep{malloci2007}, as well as IP$^\mathrm{exp}$ and AE$^{(Z+1)+}$ from experiments \citep[][and this work]{clar1981, hager1988, zhen2016a}. We compared this data set with two analytical descriptions. 
Both start from a classical model of the energy it takes to remove one electronic charge from a small, conducting particle. \citet{weingartner2001a} considered conducting spheres, and added an empirical correction term to account for both quantum effects and PAH geometry, which is planar and not spherical. This additional term was determined to fit a set of data on first and second ionization potentials of PAHs and led to 

\begin{equation}\label{eq:IP}
\mathrm{IP}^{(Z+1)+}_{WD} = W + \frac{1}{4 \pi \varepsilon_0} \left[ \left( Z+\frac{1}{2} \right) \frac{e^2}{a} + (Z+2) \frac{e^2}{a} \frac{0.03\,\text{nm}}{a} \right] \frac{1\,\mathrm{C}}{e},
\end{equation}
where $Z$ is the ion charge, $\varepsilon_0$ is the vacuum permittivity in $\mathrm{F \cdot nm^{-1}}$, $e$ is the elementary charge in C, $W$ is the work function of bulk graphite, $W=4.4\,\mathrm{eV}$, and $a$, the effective radius in nm, is proportional to $N_\mathrm{C}$ via the relation

\begin{equation}
    a = \sqrt[3]{\frac{N_\mathrm{C}}{468}} .
    \label{eq:radius}
\end{equation}

In Eq.~(\ref{eq:IP}), the first term in square brackets corresponds to the classical conducting sphere model, the second term is the empirical correction. For the data set reported in Fig.~\ref{fig:IP_AE}, we found that, instead of $W=4.4\,\mathrm{eV}$, a value of $W=3.9\,\mathrm{eV}$ fits the data better (black curves in Fig.~\ref{fig:IP_AE}). A satisfactory model is obtained for all $Z$ values considered ($Z=0,1,2$). The empirical formula which was adjusted for $Z=0,1$ appears also to be appropriate for $Z=2$. The adjustment that we made on the $W$ value corresponds to a vertical shift and somehow depends on the considered data set. For instance, we can see that our reported values for AE$^{(Z+1)+}$ are systematically slightly above the DFT values (cf. Table~\ref{table:IP}). Still, our derived value for $W$ appears in line with the values of about $4.0\,\mathrm{eV}$, which were calculated for similarly sized PAHs by \citet{kvashnin2013}.

The second formalism to describe IP$^{(Z+1)+}$ is given by \cite{bakes1994a} who considered a thin conducting disk instead of a sphere yielding

\begin{equation}\label{eq:IP_BT}
\mathrm{IP}^{(Z+1)+}_{BT} = W + \left( Z+\frac{1}{2} \right) \frac{25.1}{\sqrt{N_\mathrm{C}}}.
\end{equation}

The gray curves in Fig.~\ref{fig:IP_AE} have been obtained from Eq.~(\ref{eq:IP_BT}) \citep{bakes1994a} and using $W=3.9\,\mathrm{eV}$, as derived above. The curve provides a very satisfactory description of the data for $Z=0$, but tends to increasingly fail for higher $Z$ values. This trend was already noticed by \cite{weingartner2001a} and we can see that the discrepancy even increases for $Z=2$. Tuning the value of W does not change the shape of the curves and this emphasizes the need to include quantum effects in the estimation of IP$^{(Z+1)+}$.

We can see from Fig.~\ref{fig:IP_AE} and Eq.~(\ref{eq:IP}) that a fraction of the photons absorbed in \ion{H}{I} regions can induce ionization of PAH cations. Taking the absorption and ionization cross sections shown in Fig.~\ref{fig:cs} and considering the radiation field of the prototypical NGC\,7023 NW PDR \citep{joblin2018}, we can estimate that typically one photon over three absorbed in the $[10-13.6]\,\mathrm{eV}$ range by PAH cations with $N_\mathrm{C}=30-36$ will lead to ionization. The fraction of ionizing events will increase with increasing molecular size as the ionization potential shifts to lower energies. It reaches 0.5 for C$_{48}$H$_{20}^+$. We also note that the formation of C$_{60}^{2+}$ will be more difficult to achieve than that of a PAH$^{2+}$ of similar size, since the corresponding cations have relatively similar absorption cross sections but the value of $\mathrm{AE}^{2+}$ for C$_{60}^+$ is significantly higher, $(10.5 \pm 0.1)\,\mathrm{eV}$ \citep{douix2017}, compared to 8.7\,eV for a PAH$^+$ with $N_\mathrm{C}=60$ (see Fig.~\ref{fig:IP_AE}).

\subsection{Photoionization yield}\label{sec:PIY}

\begin{figure}[htbp]
    \resizebox{\hsize}{!}{
    \includegraphics{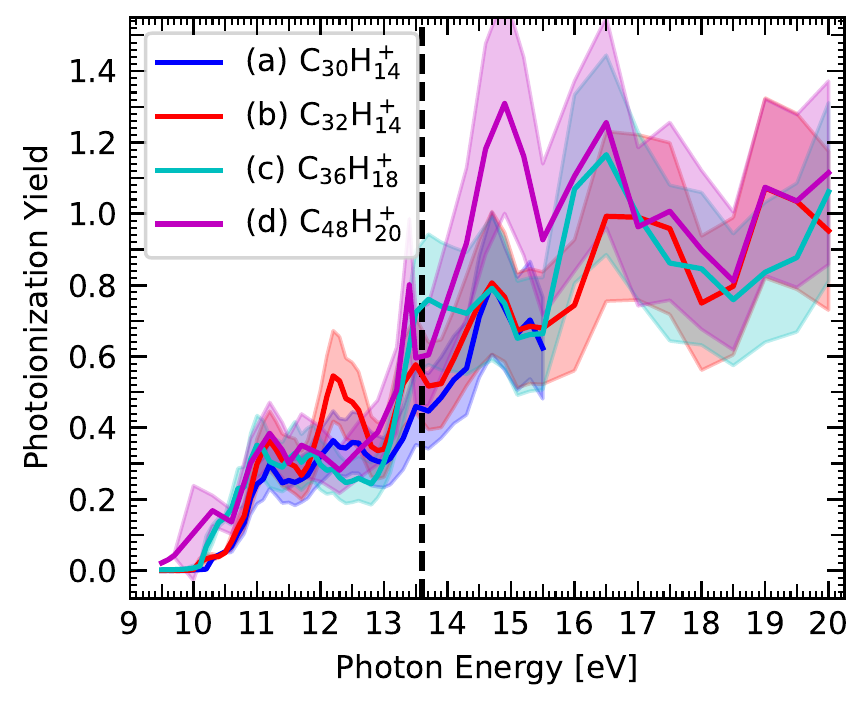}}
    \caption{Photoionization yields of the studied cations derived from the ionization and absorption cross sections (see Fig.~\ref{fig:cs}) and scaled as explained in Sect.~\ref{sec:PIY}. The dashed vertical line marks the $13.6\,\mathrm{eV}$ photon energy cut-off for \ion{H}{I} regions.}
    \label{fig:yields}
\end{figure}

Photoionization yields of PAH cations were derived for all studied species by dividing the experimental photoionization cross section, $\sigma^\mathrm{C}_\mathrm{I}$, by the theoretical photoabsorption cross section, $\sigma^\mathrm{C}_\mathrm{abs,\,theo}$.
In Sect.~\ref{sec:PIcross}, we discussed the precision of both the experimental and theoretical cross sections. This can impact the photoionization yields. At energies below 14\,eV, the presence of bands in $\sigma^\mathrm{C}_\mathrm{abs,\,theo}$, which are not present in $\sigma^\mathrm{C}_\mathrm{I}$, can induce spectral features in the photoionization yields (e.g. the 12\,eV peak obtained for C$_{32}$H$_{14}^+$), which are as precise as the calculated spectrum. Still, Fig.~\ref{fig:yields} shows that the photoionization yields display comparable features for the studied molecules, with a rise starting at the ionization thresholds, AE$^{2+}$, the plateau in the 11.3 to 12.9\,eV range followed by another rise to reach the maximum value. There is some uncertainty on this maximum value because of the unknown contribution from $\sigma^\mathrm{'}$ (cf. Sect.~\ref{sec:PIcross}). In the following, we made the hypothesis that the contribution of $\sigma^\mathrm{'}$ at high energies (20\,eV) is minor and that the photoionization yields are limited by the photoionization BR, which never reaches unity as shown in Fig.~\ref{fig:br}. The mean values at high energies of the photoionization yields were thus scaled using the ionization BR at 20\,eV. The resulting curves are presented in Fig.~\ref{fig:yields}.

Data on the photoionization yields of neutral PAHs have been previously derived from experimental studies performed by \cite{verstraete1990} and \cite{jochims1996}.
The latter authors have proposed a \emph{rule of thumb} to facilitate the implementation of this yield into models. This consists of a linear function of the photon energy with dependence on the ionization potential. On the basis of our results, we propose to use a similar approach to describe the evolution of the photoionization yield of PAH cations with molecular size. The resulting function, $Y_\mathrm{ion}^+$, is based on the above described ionization regimes which occur in different energy ranges (values in eV) as
\begin{equation}
    Y_\mathrm{ion}^+[N_\mathrm{C}](h\nu)=
    \left\{
    \begin{array}{@{}l@{}c@{}l@{}}
        \begin{array}{@{}l@{}}
        0\\
        \frac{\alpha}{11.3-\mathrm{IP}^{2+}}(h \nu -\mathrm{IP}^{2+})\\
        \alpha\\
        \frac{\beta(N_\mathrm{C})-\alpha}{2.1}\,(h \nu - 12.9) + \alpha\\
        \beta(N_\mathrm{C})
        \end{array} & 
        \text{for} & 
        \begin{array}{r@{}}
        h \nu < \mathrm{IP}^{2+}\; \\ 
        \mathrm{IP}^{2+} \leq h \nu < 11.3\; \\
        11.3 \leq h \nu < 12.9\; \\
        12.9 \leq h \nu < 15.0\; \\
        h \nu \geq 15.0,
        \end{array}
        \label{eq:yield}
    \end{array}
    \right.
\end{equation}
where $\alpha = 0.3$ is the value of the plateau and $\beta$ depends on $N_\mathrm{C}$ with
\begin{equation}
    \beta(N_\mathrm{C})=
    \left\{
    \begin{array}{@{}lcl@{}}
        \begin{array}{@{}l@{}}
        0.59 + 8.1 \cdot 10^{-3} N_\mathrm{C}\\ 
        1 
        \end{array} & \text{for} &  \begin{array}{r@{}}
        32 \leq N_\mathrm{C} < 50\; \\ 
        N_\mathrm{C} \geq 50.
        \end{array} 
    \end{array}
    \right.
    \label{eq:beta}
\end{equation}
The reported $\beta$ values represent the values at 20\,eV of the ionization BR (Fig.~\ref{fig:br}). They can be considered as maximum values since they neglect a possible contribution of $\sigma^\mathrm{'}$ to the photoabsorption cross section as discussed above. These values were found to increase linearly with size for the studied size range with the dependence given by Eq.~(\ref{eq:beta}). Extrapolation to larger sizes leads to a $\beta$ value of 1 for $N_\mathrm{C} \geq 50$. This trend differs from the case of neutral PAHs for which \citet{jochims1996} concluded that $\beta=1$ is independent of size, in agreement with previous measurements by \citet{verstraete1990}. 

\begin{figure}[htbp]
    \resizebox{\hsize}{!}{
    \includegraphics{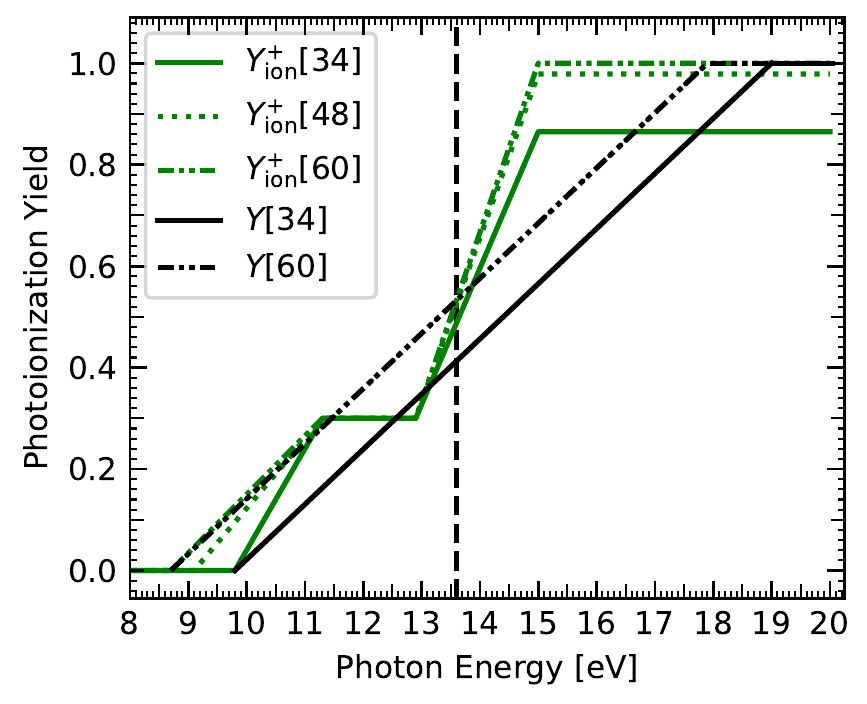}}
    \caption{Photoionization yields, $Y^+_\mathrm{ion}[N_\mathrm{C}](h \nu)$, calculated from Eqs.~(\ref{eq:yield}) and (\ref{eq:beta}) for PAH cations with $N_\mathrm{C}=34$, 48, and 60 atoms. For comparison, the photoionization yields, $Y[N_\mathrm{C}](h \nu)$, for $N_\mathrm{C}=34$ and 60, are displayed. They were estimated using the Eqs.~(4) and (5) from \citet{jochims1996} adapted for neutral PAHs but taking into account the shift of the ionization potential to IP$^{2+}$, which is relevant for cations. The dashed vertical line marks the $13.6\,\mathrm{eV}$ photon energy cut-off for \ion{H}{I} regions.}
    \label{fig:yield}
\end{figure}

Figure~\ref{fig:yield} displays examples of $Y_\mathrm{ion}^+[N_\mathrm{C}](h \nu)$ which were calculated from Eqs.~(\ref{eq:yield}) and (\ref{eq:beta}), illustrating the variability of $Y_\mathrm{ion}^+[N_\mathrm{C}](h \nu)$ with molecular size. No significant variation of this yield is expected for PAH cations with $N_\mathrm{C} \geq 50$. In their PAH evolution model, \cite{andrews2016} have considered the yield of PAH cations based on the recipe given by \cite{jochims1996} for neutrals but taking into account the appropriate photoionization potential for cations, i.e., values of IP$^{2+}$. To illustrate the impact that this approximation may have on the model results, we report in Fig.~\ref{fig:yield} these estimated yields and compare them with our recommended yields by integrating from IP$^{2+}$ to 13.6\,eV. We found that for the medium-sized PAHs, as represented by $N_\mathrm{C}=34$, our integrated yield is larger by 19\,\% compared to the previously available one, whereas for large PAHs, as represented by $N_\mathrm{C}=60$, it is smaller by 14\,\%. These simple estimates are however not conclusive and models have to be run to evaluate the impact on the ionization of the PAH population in specific environments.

\section{Conclusion}\label{sec:con}

We have studied the interaction of trapped PAH cations with VUV photons in the range of 9 to 20\,eV, covering also photon energies present in H\textsc{ii} regions and ionization fronts. Our experimental results provide a wealth of information on both ionization and fragmentation processes. The present article is focused on the detailed analysis of ionization, whereas fragmentation will be the subject of a future, dedicated work. Our initial goal was to explore the properties of large species for $N_\mathrm{C}$ up to about 80 atoms. However we could only achieve measurements on molecular sizes from 30 to 48 carbon atoms due to the very low solubility of large PAHs. Still, studies in this range allow us to access the major trends in the ionization properties of PAH cations due to a molecular size increase. We found that
\begin{itemize}
    \item[(i)] below 13.6\,eV, the formation of a hot ion with subsequent (radiative) cooling is the major relaxation channel, followed by ionization whose yield reaches about 0.5 at 13.6\,eV. From a molecular physics point of view, the yield comprises an interesting plateau at a value of 0.3 that extends over the energy range from 11.3 to 12.9\,eV. This plateau reveals a spectral range in which there is a strong competition between electronic and nuclear states. It would be interesting to investigate the dynamics of the relaxation of excited electronic states in this range using fs pump-probe experiments \citep{marciniak2015}.
    \item[(ii)] contrary to previous studies on neutrals, we could not observe that the photoionization yield reaches a value of 1 at high energies. At 20\,eV, some dissociation is observed for all studied PAH cations, implying that the maximum of the yield cannot be larger than the branching ratio between ionization and dissociation, which increases with molecular size and reaches 0.98 for the largest studied ion, C$_{48}$H$_{20}^+$. In addition, we have not included a possible contribution in the photoabsorption events of the formation of a hot ion that would subsequently relax by radiative cooling in isolated conditions. This contribution would further lower the values of the photoionization yield. We have no explanation for the difference observed between neutrals and cations. Whether this is due to a change in their respective properties or the fact that experiments like ours using ion trapping are more sensitive to quantify this effect than experiments carried out on neutrals with different techniques, is out of our reach and would be interesting to further investigate.
\end{itemize}
Concerning astrophysical applications, we provide recipes to determine both the ionization potential and the photoionization yield of PAH cations as a function of their molecular size, which can be extended to larger sizes (typically $N_\mathrm{C}=100$). This yield can be combined with photoabsorption cross sections that are readily available from calculations using TD-DFT. All this molecular data can be used in models that describe the chemical evolution of PAHs in astrophysical environments. The range of photon energy we studied makes it possible to tackle the evolution of PAHs in extreme astronomical environments such as \ion{H}{II} regions and ionization fronts. Observations of PAHs in these environments have been so far scarce due to their technical difficulty, but will become much more accessible thanks to the unique capabilities of the forthcoming James Webb Space Telescope (JWST).

As another example, the cavity around the star in NGC\,7023 is expected to be an environment in which large PAH$^{+}$ and PAH$^{2+}$ are present \citep{andrews2016, croiset16}. The presence of dications is expected to impact both the heating of the gas by photoelectric effect and the AIB emission. Some first IR action spectra of large PAH cations and dications have been recorded by \cite{zhen2017, zhen2018}. They provide encouraging results about large ionized PAHs being good candidates for carrying the AIBs. It is still not clear though if the spectral differences between cations and dications will be sufficient to differentiate both charge states in the observations. Still, we can predict that a detailed modeling approach combined with the wealth of spectral and spatial information, which will be delivered soon by the JWST, will be able to highlight the charge evolution of the PAH population and its impact on the physics and chemistry of PDRs.

\begin{acknowledgements}
This paper is dedicated to Sydney Leach, a great scientist who has been a pioneer in laboratory astrophysics and a major source of inspiration for the whole field. We are grateful to the staff from SOLEIL for the smooth running of the facility. redWe also wish to acknowledge the insightful comments of the referee on an earlier version of this paper. We acknowledge funding from the European Research Council under the European Union's Seventh Framework Programme ERC-2013-SyG, Grant Agreement no. 610256, NANOCOSMOS. This work was also supported by the Agence Nationale de la Recherche (France), under project number ANR-08-BLAN-0065. G.~W. thanks the European Union (EU) for support under the Horizon 2020 framework for the Marie Sk\l odowska-Curie action EUROPAH, Grant Agreement no. 722346. S.~Q. and D.~P. acknowledge financial support from the Spanish Agencia Estatal de Investigaci\'{o}n (MAT2016-78293-C6-3-R; AEI/FEDER, UE), Xunta de Galicia (Centro Singular de Investigaci\'{o}n de Galicia accreditation 2016--2019, ED431G/09), the European Regional Development Fund-ERDF, and the European FET-OPEN project SPRING, Grant Agreement no. 863098. Finally, this project was granted access to the HPC resources at the CALMIP supercomputing centre under project P20027.
\end{acknowledgements}

%%% REFERENCES
\bibliographystyle{aa}
\bibliography{VUVPAH_ion}

\begin{thebibliography}{39}
\expandafter\ifx\csname natexlab\endcsname\relax\def\natexlab#1{#1}\fi

\bibitem[{{Andrews} {et~al.}(2015){Andrews}, {Boersma}, {Werner}, {Livingston},
  {Allamandola}, \& {Tielens}}]{andrews2015}
{Andrews}, H., {Boersma}, C., {Werner}, M.~W., {et~al.} 2015, \apj, 807, 99

\bibitem[{Andrews {et~al.}(2016)Andrews, Candian, \& Tielens}]{andrews2016}
Andrews, H., Candian, A., \& Tielens, A. G. G.~M. 2016, \aap, 595, A23

\bibitem[{Baiardi {et~al.}(2017)Baiardi, Paoloni, Barone, Zakrzevski, \&
  Ortiz}]{baiardi2017}
Baiardi, A., Paoloni, L., Barone, V., Zakrzevski, V.~G., \& Ortiz, J.~V. 2017,
  J. Chem. Theory Comput., 13, 3120

\bibitem[{Bakes \& Tielens(1994)}]{bakes1994}
Bakes, E. L.~O. \& Tielens, A. G. G.~M. 1994, ASPC, 58, 412

\bibitem[{{Bakes} \& {Tielens}(1994)}]{bakes1994a}
{Bakes}, E.~L.~O. \& {Tielens}, A.~G.~G.~M. 1994, \apj, 427, 822

\bibitem[{Bakes {et~al.}(2001)Bakes, Tielens, \& Charles
  W.~Bauschlicher}]{bakes2001a}
Bakes, E. L.~O., Tielens, A. G. G.~M., \& Charles W.~Bauschlicher, J. 2001,
  \apj, 556, 501

\bibitem[{Bréchignac {et~al.}(2014)Bréchignac, Garcia, Falvo, Joblin, Kokkin,
  Bonnamy, Parneix, Pino, Pirali, Mulas, \& Nahon}]{brechignac2014}
Bréchignac, P., Garcia, G.~A., Falvo, C., {et~al.} 2014, \jcp, 141, 164325

\bibitem[{Cataldo {et~al.}(2011)Cataldo, Ursini, Angelini, \&
  {Iglesias-Groth}}]{cataldo2011}
Cataldo, F., Ursini, O., Angelini, G., \& {Iglesias-Groth}, S. 2011, Fuller.
  Nanotub. Car.~N., 19, 713

\bibitem[{Clar {et~al.}(1981)Clar, Robertson, Schloegl, \& Schmidt}]{clar1981}
Clar, E., Robertson, J.~M., Schloegl, R., \& Schmidt, W. 1981, J. Am. Chem.
  Soc., 103, 1320

\bibitem[{{Compi{\`e}gne} {et~al.}(2007){Compi{\`e}gne}, {Abergel},
  {Verstraete}, {Reach}, {Habart}, {Smith}, {Boulanger}, \&
  {Joblin}}]{compiegne2007}
{Compi{\`e}gne}, M., {Abergel}, A., {Verstraete}, L., {et~al.} 2007, \aap, 471,
  205

\bibitem[{Croiset {et~al.}(2016)Croiset, Candian, Bern\'e, \&
  Tielens}]{croiset16}
Croiset, B.~A., Candian, A., Bern\'e, O., \& Tielens, A. G. G.~M. 2016, \aap,
  590, A26

\bibitem[{Deleuze {et~al.}(2001)Deleuze, Trofimov, \& Cederbaum}]{deleuze2001}
Deleuze, M.~S., Trofimov, A.~B., \& Cederbaum, L.~S. 2001, J.~Chem.~Phys., 115,
  5859

\bibitem[{Douix {et~al.}(2017)Douix, Duflot, Cubaynes, Bizau, \&
  Giuliani}]{douix2017}
Douix, S., Duflot, D., Cubaynes, D., Bizau, J.-M., \& Giuliani, A. 2017,
  J.~Phys.~Chem.~Lett., 8, 7

\bibitem[{Giuliani {et~al.}(2012)Giuliani, Giorgetta, Ricaud, Jamme, Rouam,
  Wien, Laprévote, \& Réfrégiers}]{giuliani2012}
Giuliani, A., Giorgetta, J.-L., Ricaud, J.-P., {et~al.} 2012, Nucl. Instrum.
  Methods Phys. Res. B, 279, 114

\bibitem[{Hager \& Wallace(1988)}]{hager1988}
Hager, J.~W. \& Wallace, S.~C. 1988, Anal. Chem., 60, 5

\bibitem[{{Joblin} {et~al.}(2018){Joblin}, {Bron}, {Pinto}, {Pilleri}, {Le
  Petit}, {Gerin}, {Le Bourlot}, {Fuente}, {Berne}, {Goicoechea}, {Habart},
  {K{\"o}hler}, {Teyssier}, {Nagy}, {Montillaud}, {Vastel}, {Cernicharo},
  {R{\"o}llig}, {Ossenkopf-Okada}, \& {Bergin}}]{joblin2018}
{Joblin}, C., {Bron}, E., {Pinto}, C., {et~al.} 2018, \aap, 615, A129

\bibitem[{Jochims {et~al.}(1996)Jochims, Baumgaertel, \& Leach}]{jochims1996}
Jochims, H.~W., Baumgaertel, H., \& Leach, S. 1996, \aap, 314, 1003

\bibitem[{Kvashnin {et~al.}(2013)Kvashnin, Sorokin, Br{\"u}ning, \&
  Chernozatonskii}]{kvashnin2013}
Kvashnin, D.~G., Sorokin, P.~B., Br{\"u}ning, J.~W., \& Chernozatonskii, L.~A.
  2013, Appl.~Phys.~Lett., 102, 183112

\bibitem[{Le~Page {et~al.}(2003)Le~Page, Snow, \& Bierbaum}]{lepage2003}
Le~Page, V., Snow, T.~P., \& Bierbaum, V.~M. 2003, \apj, 584, 316

\bibitem[{Malloci {et~al.}(2007{\natexlab{a}})Malloci, Joblin, \&
  Mulas}]{malloci2007db}
Malloci, G., Joblin, C., \& Mulas, G. 2007{\natexlab{a}}, Chem.~Phys., 332, 353

\bibitem[{Malloci {et~al.}(2007{\natexlab{b}})Malloci, Joblin, \&
  Mulas}]{malloci2007}
Malloci, G., Joblin, C., \& Mulas, G. 2007{\natexlab{b}}, \aap, 462, 627

\bibitem[{Malloci {et~al.}(2004)Malloci, Mulas, \& Joblin}]{malloci2004}
Malloci, G., Mulas, G., \& Joblin, C. 2004, \aap, 426, 105

\bibitem[{Marciniak {et~al.}(2015)Marciniak, Despr{\'e}, Barillot, Rouz{\'e}e,
  Galbraith, Klei, Yang, Smeenk, Loriot, Reddy, Tielens, Mahapatra, Kuleff,
  Vrakking, \& L{\'e}pine}]{marciniak2015}
Marciniak, A., Despr{\'e}, V., Barillot, T., {et~al.} 2015, Nat. Commun., 6, 1

\bibitem[{{Matsuzawa} {et~al.}(2001){Matsuzawa}, {Ishitani}, {Dixon}, \&
  {Uda}}]{matsukawa2001}
{Matsuzawa}, N.~N., {Ishitani}, A., {Dixon}, D.~A., \& {Uda}, T. 2001, J. Phys.
  Chem. A, 105, 4953

\bibitem[{Milosavljevi{\'c} {et~al.}(2012)Milosavljevi{\'c}, Nicolas, Gil,
  Canon, R{\'e}fr{\'e}giers, Nahon, \& Giuliani}]{milosavljevic2012}
Milosavljevi{\'c}, A.~R., Nicolas, C., Gil, J.-F., {et~al.} 2012, J.
  Synchrotron Radiat., 19, 174

\bibitem[{Montillaud {et~al.}(2013)Montillaud, Joblin, \&
  Toublanc}]{montillaud2013}
Montillaud, J., Joblin, C., \& Toublanc, D. 2013, \aap, 552, A15

\bibitem[{Nahon {et~al.}(2012)Nahon, {de Oliveira}, Garcia, Gil, Pilette,
  Marcouill{\'e}, Lagarde, \& Polack}]{nahon2012}
Nahon, L., {de Oliveira}, N., Garcia, G.~A., {et~al.} 2012, J. Synchrotron
  Radiat., 19, 508

\bibitem[{Tancogne-Dejean {et~al.}(2020)Tancogne-Dejean, {Oliveira}, {Andrade},
  {Appel}, {Borca}, {Le Breton}, {Buchholz}, {Castro}, {Correa}, {De
  Giovannini}, {Delgado}, {Eich}, {Flick}, {Gil}, {Gomez}, {Helbig},
  {H√ºbener}, {Jest√§dt}, {Jornet-Somoza}, {Larsen}, {Lebedeva},
  {L√ºders}, {Marques}, {Ohlmann}, {Pipolo}, {Rampp}, {Rozzi}, {Strubbe},
  {Sato}, {Sch√§fer}, {Theophilou}, {Welden}, \& A.~{Rubio}}]{octopus2020}
Tancogne-Dejean, N., {Oliveira}, M. J.~T., {Andrade}, X., {et~al.} 2020, \jcp,
  152, 124119

\bibitem[{Tielens(2005)}]{tielens2005}
Tielens, A. G. G.~M. 2005, The Physics and Chemistry of the Interstellar Medium
  (Cambridge University Press)

\bibitem[{Verstraete {et~al.}(1990)Verstraete, Leger, D'Hendecourt, Defourneau,
  \& Dutuit}]{verstraete1990}
Verstraete, L., Leger, A., D'Hendecourt, L., Defourneau, D., \& Dutuit, O.
  1990, \aap, 237, 436

\bibitem[{{Vicente} {et~al.}(2013){Vicente}, {Bern{\'e}}, {Tielens},
  {Hu{\'e}lamo}, {Pantin}, {Kamp}, \& {Carmona}}]{vicente2013}
{Vicente}, S., {Bern{\'e}}, O., {Tielens}, A.~G.~G.~M., {et~al.} 2013, \apjl,
  765, L38

\bibitem[{Visser {et~al.}(2007)Visser, Geers, Dullemond, Augereau, Pontoppidan,
  \& van Dishoeck}]{visser2007}
Visser, R., Geers, V.~C., Dullemond, C.~P., {et~al.} 2007, \aap, 466, 229

\bibitem[{Weingartner \& Draine(2001{\natexlab{a}})}]{weingartner2001a}
Weingartner, J.~C. \& Draine, B.~T. 2001{\natexlab{a}}, \apj, 548, 296

\bibitem[{Weingartner \& Draine(2001{\natexlab{b}})}]{weingartner2001b}
Weingartner, J.~C. \& Draine, B.~T. 2001{\natexlab{b}}, \apjs, 134, 263

\bibitem[{{Yabana} \& {Bertsch}(1996)}]{yabana1996}
{Yabana}, K. \& {Bertsch}, G.~F. 1996, Phys. Rev. B, 54, 4484

\bibitem[{{Zhen} {et~al.}(2018){Zhen}, {Candian}, {Castellanos}, {Bouwman},
  {Linnartz}, \& {Tielens}}]{zhen2018}
{Zhen}, J., {Candian}, A., {Castellanos}, P., {et~al.} 2018, \apj, 854, 27

\bibitem[{{Zhen} {et~al.}(2017){Zhen}, {Castellanos}, {Bouwman}, {Linnartz}, \&
  {Tielens}}]{zhen2017}
{Zhen}, J., {Castellanos}, P., {Bouwman}, J., {Linnartz}, H., \& {Tielens}, A.
  G.~G.~M. 2017, \apj, 836, 28

\bibitem[{Zhen {et~al.}(2015)Zhen, Castellanos, Paardekooper, Ligterink,
  Linnartz, Nahon, Joblin, \& Tielens}]{zhen2015}
Zhen, J., Castellanos, P., Paardekooper, D.~M., {et~al.} 2015, \apj, 804, L7

\bibitem[{Zhen {et~al.}(2016)Zhen, {Rodriguez Castillo}, Joblin, Mulas, Sabbah,
  Giuliani, Nahon, Martin, Champeaux, \& Mayer}]{zhen2016a}
Zhen, J., {Rodriguez Castillo}, S., Joblin, C., {et~al.} 2016, \apj, 822, 113

\end{thebibliography}

\begin{appendix}

\section{Synthesis of dicoronylene, C$_\mathsf{48}$H$_\mathsf{20}$}\label{app:dico}

Dicoronylene was synthesized following the procedure reported by \citet{cataldo2011}. A mixture of coronene ($0.060\,\mathrm{g}$, $0.199\,\mathrm{mmol}$), AlCl$_3$ ($4.20\,\mathrm{g}$, $31.2\,\mathrm{mmol}$), NaCl ($0.817\,\mathrm{g}$, $14.0\,\mathrm{mmol}$), and CuCl ($0.043\,\mathrm{g}$, $0.440\,\mathrm{mmol}$) was stirred at $195\,^\circ\mathrm{C}$ for $2\,\mathrm{h}$. Then, aqueous solution of HCl ($10\,\%$, $20\,\mathrm{ml}$) was slowly added resulting in the precipitation of a red solid. This solid was filtrated and washed with aqueous solution of HCl ($10\,\%$, $20\,\mathrm{ml}$), hot water ($20\,\mathrm{ml}$) and acetone ($3 \times 20\,\mathrm{ml}$). The resulting solid was dried to obtain dicoronylene as a red solid ($54\,\mathrm{mg}$, $40\,\%$).

\begin{figure}[htbp]
    \resizebox{\hsize}{!}{
    \includegraphics{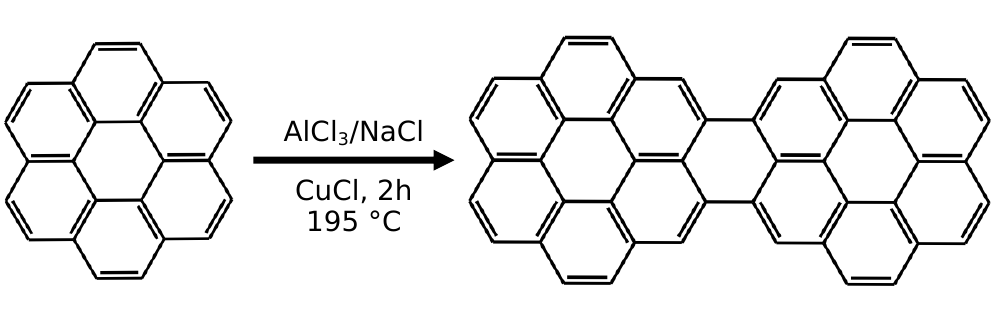}}
    \caption{Schematic of the synthesis of dicoronylene, C$_{48}$H$_{20}$, from coronene, C$_{24}$H$_{12}$.}
    \label{fig:dico}
\end{figure}

\section{Action spectra scaling procedure}\label{app:actionspec}
From the recorded mass spectra, peak intensities of parent ions and their photoproducts can be deduced. The secondary ions produced upon VUV irradiation consist of the dication with peak intensity $S_\mathrm{I}$, and the photofragments with summed peak intensity $S_\mathrm{F}$, including the H and 2H/H$_2$ loss channels. Due to detector characteristics, doubly ionized molecules are detected more efficiently than singly ionized molecules. Therefore, the peak intensities derived from the mass spectra have to be scaled by the detector gain efficiency, $\varepsilon$, to retrieve values that scale with abundances. Thermo Scientific\texttrademark\ provides $\varepsilon_+=0.29$ for parent ions and fragments, $\varepsilon_{2+}=0.42$ for dications and a value of 0.54 for trications. There is some gain change with mass but this is a minor correction for the range of studied masses.

The total number of ions in the trap, $P_0$ (in uncalibrated values), can then be calculated with
\begin{equation}
    \label{eq:p0}
    P_0 = \frac{P_t}{\varepsilon_+} + \frac{S_\mathrm{F}}{\varepsilon_+} + \frac{S_\mathrm{I}}{\varepsilon_{2+}},
\end{equation} 
where $\frac{P_t}{\varepsilon_+}$ is the number of parent ions after irradiation time, $t$. Building the action spectra requires to derive normalized photoproduct intensities for $S_\mathrm{I}$ and $S_\mathrm{F}$, which can be obtained by first dividing them by $P_0$ and then correcting for the variation of the photon flux, $\Phi(\nu)$. Indeed the latter evolved in energy due to spectral shape variations and changes made in the irradiation time, $t$, and monochromator exit slit width, $s$, so that the total photoproduct intensity remains smaller than \textit{ca.} 12\,\% of $P_t$, as seen in the mass spectra. This led to
\begin{equation}
    S^\mathrm{norm}_\mathrm{I, F}(\nu) = \frac{S_\mathrm{I, F}(\nu)}{\Phi_\mathrm{norm}(\nu) \, t(\nu) \, s(\nu)},
\end{equation}
where $\Phi_\mathrm{norm}(\nu)$ is normalized to be 1 at its maximum at $9.5\,\mathrm{eV}$. Note that the obtained intensities are in arbitrary units (see Fig.~\ref{fig:ri}) and not in percentage of the total number of ions because of the scaling by the relative photon flux.

\section{Photoproduct cross sections scaling procedure}\label{app:cs}
The absorption of a photon by the parent ion leads to different relaxation channels. Considering monochromatic radiation, the photoabsorption cross section, $\sigma_\mathrm{abs}$, can therefore be decomposed into the sum of the cross sections for each relaxation channel as 
\begin{equation}
    \label{eq:sabs}
    \sigma_\mathrm{abs} = \sigma_\mathrm{I+F} + \sigma^\mathrm{'},
\end{equation}
where $\sigma_\mathrm{I+F}$ is the cross section leading to the production of the secondary products with intensities $S_\mathrm{I}$ and $S_\mathrm{F}$ (see Appendix~\ref{app:actionspec}), and $\sigma^\mathrm{'}$ is the cross section for the creation of a hot ion that will relax its internal energy by radiative cooling and/or collisions with buffer gas (He) in our experiment. These processes cannot be traced in our experiment and only $\sigma_\mathrm{I+F}$ can be estimated following
\begin{equation}
    \label{eq:cs1}
    \sigma_\mathrm{I+F} = \frac{\gamma}{\Phi t} \ln{\left( \frac{\varepsilon_+ P_0}{P_t} \right)},
\end{equation}
with the photon flux, $\Phi$, in $\mathrm{photons \, cm^{-2}s^{-1}}$, a form factor, $\gamma$, describing the overlap of the photon beam and the ion cloud, the total number of ions, $P_0$, and the number of parent ions, $\frac{P_t}{\varepsilon_+}$, after irradiation time, $t$. Plugging $P_0$ from Eq.~(\ref{eq:p0}) in Eq.~(\ref{eq:cs1}), we get
\begin{equation}
    \label{eq:cs2}
    \sigma_\mathrm{I+F} = \frac{\gamma}{\Phi t} \ln{\left(1 + \frac{S_\mathrm{F}}{P_t} + \frac{\varepsilon_+ S_\mathrm{I}}{\varepsilon_{2+} P_t}\right)}.
\end{equation}
In order to determine cross sections in absolute units, the photon flux, $\Phi$, and the form factor, $\gamma$, have to be well-known. \citet{douix2017} managed to record the absolute photoionization cross section, $\sigma^{\mathrm{C}_{60}^+}_\mathrm{I}$, for the buckminsterfullerene cation, C$_{60}^+$, by carefully measuring these parameters and applying Eq.~(\ref{eq:cs2}), where the term $\frac{S_\mathrm{F}}{P_t}$ was zero due to the non-dissociation of C$_{60}^+$\footnote{The absolute photoionization cross section of C$_{60}^+$, $\sigma^{\mathrm{C}_{60}^+}_\mathrm{I}$, can be accessed at 10.5281/zenodo.1001072}. In our experiment, we trapped  C$_{60}^+$ and recorded its dication peak under the same irradiation conditions used for our studied PAHs except dicoronylene as explained below. We could therefore derive the value of $\frac{\gamma}{\Phi t}$ and use this factor to obtain experimental values for $\sigma_\mathrm{I+F}$ from the photoproduct evolutions of our PAHs according to Eq.~(\ref{eq:cs2}). This used scaling procedure yields reasonable cross section values only above the AE$^{2+}$ of C$_{60}^+$. We bypassed this limitation by scaling the action spectra (see Sect.~\ref{sec:act} and Appendix~\ref{app:actionspec}) to the cross sections and replacing the values below the AE$^{2+}$ of C$_{60}^+$ with the values from the scaled action spectra.
Finally, to simplify the comparison between species, we divide $\sigma_\mathrm{I+F}$ of each PAH cation by its respective $N_\mathrm{C}$, yielding $\sigma^\mathrm{C}_\mathrm{I+F}$.

We note that in this calibration procedure, dicoronylene required a specific treatment. Indeed this ion was studied in different conditions since both the syringe flow rate and the monochromator exit slit width, $s$, were increased in order to get a sufficient signal. In order to correct at best for these changes, we applied corrections to the $\frac{\gamma}{\Phi t}$ factor in Eq.~(\ref{eq:cs1}) by assuming that not only the photon flux but also the beam overlap with the ion cloud scales linearly with $s$, the later is likely disputable but this is the best we could do.

\section{Error estimation}\label{app:error}
Depending on the acquisition time, a few hundred, $N$, mass spectra, $x$, are recorded for each photon energy step. Averaging these mass spectra for each photon energy yields one mean mass spectrum, $\bar{x}$, per photon energy with an absolute standard error of $\Delta \bar{x} = \frac{\sigma}{\sqrt{N}}$, where $\sigma$ is the standard deviation. The error bars, $\Delta f$, in Figs.~\ref{fig:ri} and \ref{fig:br} result from error propagation for a function, $f \to f(x_1, x_2, ..., x_n)$, according to 
\begin{equation}
     \Delta f = \sqrt{\sum_{i=1}^n \left(\frac{\partial f}{\partial x_i} \Delta x_i\right)^2}.
\end{equation}
When determining the cross sections depicted in Fig.~\ref{fig:cs} following the procedure presented in Appendix~\ref{app:cs} as well as the obtained photoionization yields shown in Figs.~\ref{fig:yields} and \ref{fig:yield}, the error on the cross section of C$_{60}^+$, which was found to be below 21\,\% by \citet{douix2017}, is used to propagate the errors.

\section{Theoretical cross section calculations}\label{app:thcalcs}
Theoretical photoabsorption cross sections were obtained using Time-Dependent Density Functional Theory (TD-DFT), with the real-time, real-space method of \citet{yabana1996}, as implemented in the \textsc{Octopus} computer code \citep{octopus2020}. Not all species in this work were considered in our previous work \citep{malloci2007db}. In addition, since we here need to use theoretical spectra up to relatively high energies ($\sim$20\,eV), we also took the chance to verify at these energies the convergence of the calculations with respect to the simulation box size and grid spacing, the real-space equivalent of more conventional basis-set convergence for e.g. Gaussian-based DFT. It is known \citep{matsukawa2001}, and mentioned in the \textsc{Octopus} documentation that, as a rule of thumb, larger simulation boxes are needed to obtain converged values for the energies of higher lying electronic excited states, and for the intensities of the transitions involving them. Moreover, when using the real-time, real-space method implemented in \textsc{Octopus}, the photoabsorption spectrum includes both transitions to discrete (bound) electronic states and to the continuum of unbound states. The former tend to converge to well-defined energies in the limit of infinite box size; the latter, instead, are artificially quantized by the boundary conditions of a finite simulation box, whose individual energies keep on changing with changing box size (and shape).

In this context, we have started a computational effort to test convergence as a function of box sizes. It led us to conclude that the spectra in the database of \citet{malloci2007db} are adequately converged for excitation energies up to $\sim$13\,eV, meaning they are suitable for astronomical modeling purposes involving photons up to the Lyman limit. We instead had to consider substantially larger simulation box sizes, and denser grid spacing, to achieve convergence up to $\sim$20\,eV, as required for this work. For our purposes, we found that a ``minimal'' simulation box can be achieved by considering the union of  8\,\AA\ radius spheres centered on each atom of the molecule, or by a single larger sphere with an equivalent volume. In addition, a grid spacing of 0.18\,\AA\ was needed to achieve convergence. All the calculations were performed on the \textsc{Olympe} supercomputer at the mesocentre CALcul en MIdi Pyrénées (CALMIP). We retained from the set of simulations two different converged spectra for each species and averaged them together in order to smooth out the contribution of free electron states and retain discrete ones.
Finally, we divided the $\sigma_\mathrm{abs,\,theo}$ of each PAH cation by its respective $N_\mathrm{C}$, yielding $\sigma_\mathrm{abs,\,theo}^\mathrm{C}$.
\end{appendix}
\end{document}